\documentclass[twoside,twocolumn,9pt]{article}
\usepackage{extsizes}
\usepackage[super,sort&compress,comma]{natbib} 
\usepackage[version=3]{mhchem}
\usepackage[left=1.5cm, right=1.5cm, top=1.785cm, bottom=2.0cm]{geometry}
\usepackage{balance}
\usepackage{mathptmx}
\usepackage{sectsty}
\usepackage{graphicx} 
\usepackage{lastpage}
\usepackage[format=plain,justification=justified,singlelinecheck=false,font={stretch=1.125,small,sf},labelfont=bf,labelsep=space]{caption}
\usepackage{float}
\usepackage{fancyhdr}
\usepackage{fnpos}
\usepackage[english]{babel}
\usepackage{array}
\usepackage{droidsans}
\usepackage{charter}
\usepackage[T1]{fontenc}
\usepackage[usenames,dvipsnames]{xcolor}
\usepackage{setspace}
\usepackage[compact]{titlesec}
\usepackage{hyperref}
\usepackage{epstopdf}%This line makes .eps figures into .pdf - please comment out if not required.
\usepackage{comment}
\usepackage{physics}
\usepackage{siunitx}

\usepackage{mhchem}
\usepackage{bm}

\newcommand{\red}[1]{#1}

\DeclareSIUnit\Molar{M}
\DeclareSIUnit\days{days}
\DeclareSIUnit\mM{\milli\Molar}
\DeclareSIUnit{\wtpercent}{\text{wt\%}}
\DeclareSIUnit{\volpercent}{\text{vol\%}}
\DeclareSIUnit\px{px}

\definecolor{cream}{RGB}{222,217,201}
\begin{document}

\pagestyle{fancy}
\thispagestyle{plain}
\fancypagestyle{plain}{
%%%HEADER%%%
\renewcommand{\headrulewidth}{0pt}
}
%%%END OF HEADER%%%

%%%PAGE SETUP - Please do not change any commands within this section%%%
\makeFNbottom
\makeatletter
\renewcommand\LARGE{\@setfontsize\LARGE{15pt}{17}}
\renewcommand\Large{\@setfontsize\Large{12pt}{14}}
\renewcommand\large{\@setfontsize\large{10pt}{12}}
\renewcommand\footnotesize{\@setfontsize\footnotesize{7pt}{10}}
\makeatother

\renewcommand{\thefootnote}{\fnsymbol{footnote}}
\renewcommand\footnoterule{\vspace*{1pt}% 
\color{cream}\hrule width 3.5in height 0.4pt \color{black}\vspace*{5pt}} 
\setcounter{secnumdepth}{5}

\makeatletter 
\renewcommand\@biblabel[1]{#1}            
\renewcommand\@makefntext[1]% 
{\noindent\makebox[0pt][r]{\@thefnmark\,}#1}
\makeatother 
\renewcommand{\figurename}{\small{Fig.}~}
\sectionfont{\sffamily\Large}
\subsectionfont{\normalsize}
\subsubsectionfont{\bf}
\setstretch{1.125} %In particular, please do not alter this line.
\setlength{\skip\footins}{0.8cm}
\setlength{\footnotesep}{0.25cm}
\setlength{\jot}{10pt}
\titlespacing*{\section}{0pt}{4pt}{4pt}
\titlespacing*{\subsection}{0pt}{15pt}{1pt}
%%%END OF PAGE SETUP%%%

%%%FOOTER%%%
\fancyfoot{}
\fancyfoot[LO,RE]{\vspace{-7.1pt}\includegraphics[height=9pt]{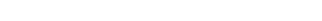}}
\fancyfoot[CO]{\vspace{-7.1pt}\hspace{13.2cm}\includegraphics{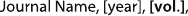}}
\fancyfoot[CE]{\vspace{-7.2pt}\hspace{-14.2cm}\includegraphics{head_foot/RF}}
\fancyfoot[RO]{\footnotesize{\sffamily{1--\pageref{LastPage} ~\textbar  \hspace{2pt}\thepage}}}
\fancyfoot[LE]{\footnotesize{\sffamily{\thepage~\textbar\hspace{3.45cm} 1--\pageref{LastPage}}}}
\fancyhead{}
\renewcommand{\headrulewidth}{0pt} 
\renewcommand{\footrulewidth}{0pt}
\setlength{\arrayrulewidth}{1pt}
\setlength{\columnsep}{6.5mm}
\setlength\bibsep{1pt}
%%%END OF FOOTER%%%

%%%FIGURE SETUP - please do not change any commands within this section%%%
\makeatletter 
\newlength{\figrulesep} 
\setlength{\figrulesep}{0.5\textfloatsep} 

\newcommand{\topfigrule}{\vspace*{-1pt}% 
\noindent{\color{cream}\rule[-\figrulesep]{\columnwidth}{1.5pt}} }

\newcommand{\botfigrule}{\vspace*{-2pt}% 
\noindent{\color{cream}\rule[\figrulesep]{\columnwidth}{1.5pt}} }

\newcommand{\dblfigrule}{\vspace*{-1pt}% 
\noindent{\color{cream}\rule[-\figrulesep]{\textwidth}{1.5pt}} }

\makeatother
%%%END OF FIGURE SETUP%%%

%%%TITLE, AUTHORS AND ABSTRACT%%%
\twocolumn[
  \begin{@twocolumnfalse}
{\includegraphics[height=30pt]{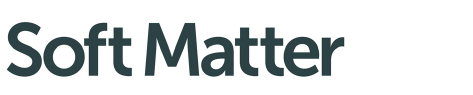}\hfill\raisebox{0pt}[0pt][0pt]{\includegraphics[height=55pt]{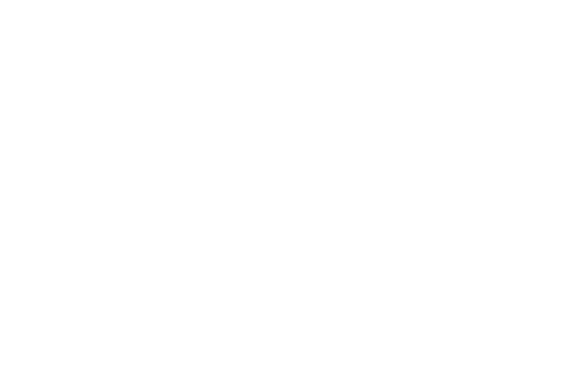}}\\[1ex]
\includegraphics[width=18.5cm]{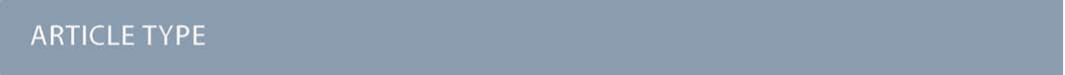}}\par
\vspace{1em}
\sffamily
\begin{tabular}{m{4.5cm} p{13.5cm} }

\includegraphics{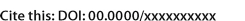} & \noindent\LARGE{\textbf{Real-space observation of salt-dependent aging in Laponite gels$^\dag$}} \\%Article title goes here instead of the text "This is the title"
\vspace{0.3cm} & \vspace{0.3cm} \\

 & \noindent\large{Shunichi Saito,$^{\ast}$\textit{$^{a}$} Sooyeon Kim,\textit{$^{b}$} Yuichi Taniguchi,\textit{$^{b,c,d}$} and Miho Yanagisawa$^{\ast}$\textit{$^{a,e,f}$}} \\%Author names go here instead of "Full name," etc.

\includegraphics{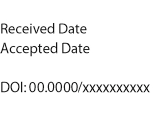} & \noindent\normalsize{
Colloidal gels gradually evolve as their structures reorganize, a process known as aging. Understanding this behavior is essential for fundamental science and practical applications such as drug delivery and tissue engineering. This study examines the aging of low-concentration Laponite suspensions with varying salt concentrations using fluorescence microscopy, scattering imaging, and particle tracking microrheology. Structural heterogeneity appeared earlier at higher salt concentrations, and the average size of aggregates decreased as the salt concentration increased further. Fourier transform analysis corroborated these trends, and scattering images showed similar results. Microrheology revealed distinct dynamics in Laponite-rich and Laponite-poor regions: the poor phase exhibited liquid-like behavior, while the rich phase exhibited gel-like properties. Further analysis suggested the presence of submicron or nanoscale structural heterogeneities within the rich phase. These findings provide insight into how aging and salt concentration shape the structure and dynamics of colloidal gels.
}
\end{tabular}

 \end{@twocolumnfalse} \vspace{0.6cm}

  ]
%%%END OF TITLE, AUTHORS AND ABSTRACT%%%

%%%FONT SETUP - please do not change any commands within this section
\renewcommand*\rmdefault{bch}\normalfont\upshape
\rmfamily
\section*{}
\vspace{-1cm}

%%%FOOTNOTES%%%

\footnotetext{\textit{$^{a}$~Graduate School of Science, The University of Tokyo, Hongo 7-3-1, Bunkyo, Tokyo 113-0033, Japan; E-mail: saito-shunichi501@g.ecc.u-tokyo.ac.jp}}
\footnotetext{\textit{$^{b}$~Institute for Integrated Cell-Material Sciences (iCeMS), Kyoto University, Yoshida-Honmachi, Sakyo-ku, Kyoto 606-850, Japan}}
\footnotetext{\textit{$^{c}$~Graduate School of Biostudies, Kyoto University, Yoshida-Honmachi, Sakyo-ku, Kyoto 606-850, Japan}}
\footnotetext{\textit{$^{d}$~Graduate School of Frontier Biosciences, Osaka University, 1-3 Yamadaoka, Suita, Osaka 565-0871, Japan}}
\footnotetext{\textit{$^{e}$~Komaba Institute for Science, Graduate School of Arts and Sciences, The University of Tokyo, 3-8-1 Komaba, Meguro, Tokyo 153-8902, Japan}}
\footnotetext{\textit{$^{f}$~Center for Complex Systems Biology, Universal Biology Institute, The University of Tokyo, Komaba 3-8-1, Meguro, Tokyo 153-8902, Japan; E-mail: myanagisawa@g.ecc.u-tokyo.ac.jp}}

%Please use \dag to cite the ESI in the main text of the article.
%If you article does not have ESI please remove the the \dag symbol from the title and the footnotetext below.

%\footnotetext{\dag~Supplementary Information available: [details of any supplementary information available should be included here]. See DOI: 10.1039/cXsm00000x/}
%additional addresses can be cited as above using the lower-case letters, c, d, e... If all authors are from the same address, no letter is required

%\footnotetext{\ddag~Additional footnotes to the title and authors can be included \textit{e.g.}\ `Present address:' or `These authors contributed equally to this work' as above using the symbols: \ddag, \textsection, and \P. Please place the appropriate symbol next to the author's name and include a \texttt{\textbackslash footnotetext} entry in the the correct place in the list.}

%%%END OF FOOTNOTES%%%

%%%MAIN TEXT%%%%

\section{Introduction}

Amorphous materials such as glasses and gels are prevalent in both natural and synthetic systems. A defining characteristic of these materials is their time-dependent evolution, called aging\cite{Berthier2011-rk}. While both glasses and gels exhibit aging, their underlying mechanisms differ fundamentally. In gels, aging involves the coarsening of structural length scales and the gradual reorganization of network structures; these features are generally absent in glasses \cite{Wang2024-kv}. Gels are widely used in applications ranging from food processing to biomedical engineering \cite{Cao2021-fa}, where aging-induced changes in structure can lead to deviations from the intended functionality. Consequently, understanding the aging behavior of gels is essential from both fundamental and applied perspectives.

Colloidal gels offer a relatively simple model system for studying aging phenomena, particularly in comparison to polymer gels, where chain entanglement complicates the dynamics \cite{Hutchinson1995-fr}. 
\red{In colloidal systems, aging originates from the progressive restriction of particle motion. In glassy states, particles become trapped in cages formed by their neighbors, leading to progressively slowed relaxation\cite{Cipelletti2005-oo}. In gels, attractive interactions form a percolated network that undergoes slow reorganization or coarsening; gelation often occurs via arrested phase separation at short-range attraction\cite{Zaccarelli2007-rs,Trappe2004-nz,Cates2004-by}. These mechanisms yield time-dependent changes in dynamics and mechanics that strengthen with aging\cite{Royall2021-jc}.}

Laponite, a synthetic, disk-shaped nanoscale clay, has received significant attention among colloidal gels. In aqueous suspension, Laponite particles exhibit an anisotropic surface charge, with negatively charged faces and positively charged rims give rise to complex electrostatic interactions \cite{Tanaka2004-ni,Ruzicka2011-zl,Suman2018-wv}. 
These interactions result in a wide range of phase behaviors.
For instance, Laponite-based nanomaterials have shown promise for applications such as sustained protein delivery\cite{Das2019-xo,Gonzalez-Pujana2024-ff}; however, aging-related changes often complicate their practical use.

Salt concentration plays a critical role in the aging behavior of Laponite suspensions by screening repulsive interactions and promoting interparticle attraction. At sufficiently high salt concentrations, this can lead to macroscopic phase separation \cite{Mourchid1998-fz,Mongondry2005-db}. Depending on the conditions, low-concentration Laponite suspensions have been interpreted as either gels  or attractive glasses resulting from phase separation, etc\cite{Nicolai2001-il,Jabbari-Farouji2008-pt,Ruzicka2006-sl}. Microrheology has revealed spatial heterogeneity in viscoelastic properties \red{\cite{Oppong2008-td,Rich2011-uj,Jabbari-Farouji2008-nl,Jabbari-Farouji2012-mw,Vyas2016-mf,Pilavtepe2018-tr}} \red{(see Table. S1 in the Supplementary Information (SI))}, while fluorescence imaging has visualized corresponding structural inhomogeneities \cite{Garcia2012-wc,Pujala2018-ks}. However, at high Laponite concentrations, structural transitions occur rapidly, hindering detailed temporal analysis. Consequently, recent systematic studies have focused on low-concentration regimes, where aging proceeds more gradually \cite{Ruzicka2011-zl}. Nevertheless, a comprehensive understanding of salt-dependent aging dynamics at low Laponite concentrations remains elusive.

In this study, we investigate the aging dynamics of low-concentration Laponite suspensions across a range of salt concentrations. By combining real-space imaging, both with and without fluorescent labeling, alongside microrheological analysis based on submicron tracer diffusion, we reveal the structural reorganization processes that occur during aging. Our results show that the observed aging behavior reflects a complex interplay between phase separation and kinetic arrest, offering new insights into the fundamental mechanisms underlying gel aging in colloidal systems.

\section{Materials and Methods}

\begin{figure}
    \centering
    \includegraphics[width=\linewidth]{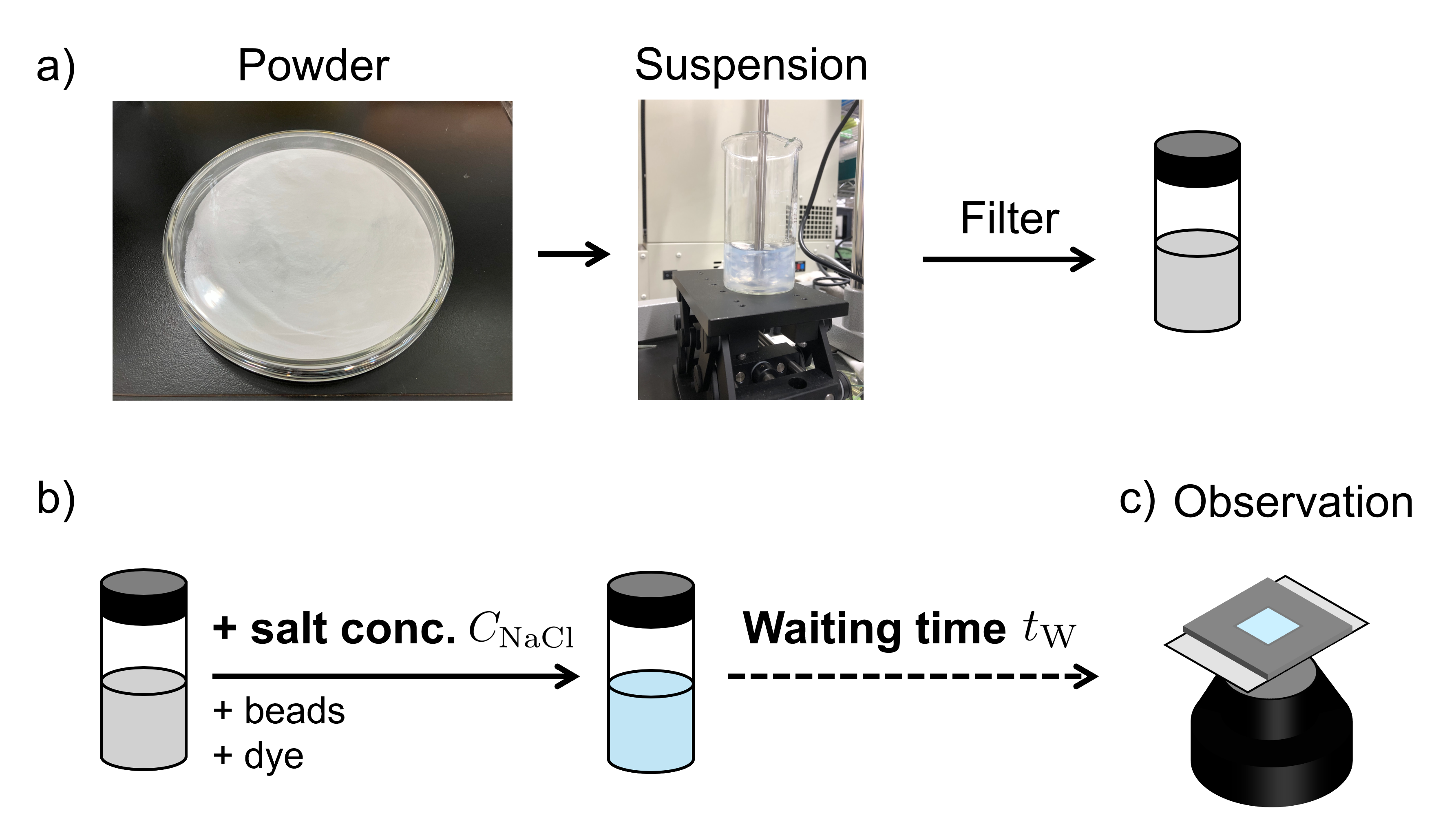}
    \caption{Preparation of Laponite suspensions.}
    \label{fig:preparation}
\end{figure}

\subsection{Materials}
Laponite RD (synthetic modified phyllosilicate) white powder was provided by BYK Chemie Japan (Tokyo, Japan). The sodium hydroxide (NaOH) solution, which served as the solvent for the Laponite suspension, was purchased from Nacalai Tesque (Kyoto, Japan). Sodium chloride (FUJIFILM Wako Pure Chemical Corporation, Tokyo, Japan) was added as a salt. For fluorescence observation, 5-carboxytetramethylrhodamine (TAMRA) (Sigma-Aldrich, Japan) was used as the fluorescent dye (molecular weight: \red{\SI{430.45}{\gram\per\mol}}). TAMRA was dissolved in 1× phosphate-buffered saline (PBS) by diluting a 10× PBS solution (304-30185, Nippon Gene Co., Ltd., Toyama, Japan) with distilled water (Invitrogen, CA, USA; Catalog no. 10977-023). The final 1x PBS composition was \SI{137}{\mM} NaCl, \SI{8.1}{\mM} \ce{Na2HPO4}, \SI{2.68}{\mM} KCl, \SI{1.47}{\mM} \ce{KH2PO4}\red{, and its ionic strength was \SI{197.1}{\mM}}. For light scattering observations, nonfluorescent polystyrene beads of various diameters were used as a control: \SI{22}{\nm} (PS02001, Bangs Laboratories, Fishers, IN), \SI{99}{\nm} (PS02004, Bangs Laboratories), \SI{300}{nm} (LB3, Sigma-Aldrich), and \SI{600}{\nm} (LB6, Sigma-Aldrich). Fluorescent polystyrene beads with a diameter of \SI{510}{\nm} (Thermo Fisher Scientific) were used as tracer beads for particle tracking analysis. Their surfaces were coated with the fluorescent dye Firefli™ Fluorescent Green.

\subsection{Suspension Preparation}
To minimize the effect of moisture absorption, Laponite powder was dried at \SI{250}{\degreeCelsius} for 18 hours using a dry heat sterilizer (WFO-420, Tokyo Rikakikai; Tokyo, Japan) and then stored in a desiccator (OH-SK, As One; Osaka, Japan) at a room temperature of around \SI{20}{\degreeCelsius} until use. To prevent chemical degradation, \SI{0.1}{\mM} NaOH aqueous solution (pH $\approx$ 9.5) was used as the solvent, and the Laponite powder was dissolved in this solution to achieve a concentration of \SI{1}{\wtpercent}. After stirring for 30 minutes, the solution was filtered through a polyvinylidene fluoride (PVDF) membrane filter (Whatman™, pore size: \SI{0.45}{\um}) to remove any undissolved aggregates (Fig. \ref{fig:preparation}a).

A salt solution of \SI{1.0}{\Molar} NaCl was first mixed with a Laponite solvent, \SI{0.1}{\mM} NaOH. The filtered \red{\SI{1}{\wtpercent}} Laponite suspension was then mixed with the solution to adjust the salt concentration. \red{This process caused a maximum decrease of \SI{0.04}{\wtpercent} in Laponite concentration, which is negligible compared to the initial \SI{1}{\wtpercent}. Therefore, the Laponite concentration will be referred to as \SI{1}{\wtpercent}, regardless of the NaCl concentration.} %Although this process slightly reduced the Laponite concentration (at most $\sim$ \SI{0.04}{\wtpercent}), the effect was considered negligible. 
This suspension was mixed using a vortex mixer.

The prepared Laponite suspension was stored in a tube with a screw cap (Maruemu, Mighty Vial, No.3). To prevent concentration changes due to evaporation, the gap between the lid and the container was sealed with Parafilm before leaving it at room temperature. This study defined the elapsed time after adding NaCl until measurement as the waiting time $t_{\mathrm{W}}$ and used it to evaluate the effects of aging.

\subsection{Fluorescence microscopic observation}
\label{fluor_observation_method}

To observe the structures formed in the Laponite suspension using fluorescence microscopy, \SI{0.1}{\mM} TAMRA dissolved in 1x PBS buffer was added to the Laponite suspension before adding salt (Fig. \ref{fig:preparation}b). The final TAMRA concentration was set to \SI{20}{\nano\Molar}. The maximum increase in salt concentration due to the addition of PBS is \SI{0.03}{\mM}, which is significantly lower than the NaOH concentration. Therefore, the impact of PBS can be ignored when estimating the salt concentration of the Laponite suspension.

The Laponite suspension containing TAMRA was enclosed in a chamber to prevent evaporation during microscopy (Fig. \ref{fig:preparation}c). The chamber was assembled with two cover glasses (C018241, C022321, Matsunami; Osaka, Japan) and an adhesive seal (SLF0201, BIO-RAD; CA, USA; \SI{9}{\mm} × \SI{9}{\mm} × \SI{0.3}{\mm}).

Fluorescence images were acquired using a confocal laser scanning microscope (IX83, FV1200; Olympus Inc., Tokyo, Japan) with a water-immersion objective lens (UPLSAPO 60XW, UPLANSAPO 20X; Olympus Inc.). TAMRA was excited with a \SI{550}{\nm} laser, and fluorescence was detected in the \SIrange{575}{675}{\nm} range using a bandpass filter. To minimize noise due to fluorescence intensity flickering, 3--5 consecutive images were taken at the same position and averaged using a Kalman filter. We confirmed that the Laponite samples did not show any obvious structural changes within the few minutes required for this imaging.

\subsection{Power spectrum of fluorescence images}
The power spectrum was calculated as follows: a two-dimensional fast Fourier transform (FFT) was applied to each image \( I(x, y) \) to obtain the Fourier coefficients \( F(k_x, k_y) \). The spatial frequencies were defined as \( k_i = n_i / N \, (i = x, y) \), where \( N \) is the image width and \( n_i \) are integer indices. The power at each frequency component was calculated as \( P(k_x, k_y) = |F(k_x, k_y)|^2 \). To obtain the radially averaged power spectrum \( P(k) \), the intensity was averaged over the magnitude of the spatial frequency \( k = \sqrt{k_x^2 + k_y^2} = \sqrt{n_x^2 + n_y^2}/N \). For efficient binning, \( \sqrt{n_x^2 + n_y^2} \) was rounded to the nearest integer, and the average was taken over all frequency components belonging to the same bin.

\subsection{Microscopic observation of Light scattering}
\label{light_scattering_method}
To visualize the aggregate structure of Laponite without fluorescent labeling, we performed light scattering measurements on a custom-built open-top light-sheet microscope, namely PISA (Planar Illumination Microscope for Single-Molecule Imaging for All-Purpose), which complements thing described above\cite{Taniguchi2024-zs}. The PISA was built on a custom microscope body with two water-immersion objective lenses, one for the illumination (54-10-7, NA = 0.66, 28.6×, Special Optics, USA) and another for the detection of scattered light (XLUMPLFLN 20XW, NA = 1.0, 20×, Evident, Japan). Two lenses were mechanically connected in an orthogonal manner and positioned below the microscope stage at a tilted angle of 33.8 degrees. For the laser source, a \SIlist[list-final-separator = {/}, list-pair-separator = {/}]{488;514.5}{\nm} Argon ion laser (Innova70-C, Coherent, Saxonburg, PA) was used. In this study, the wavelength of \SI{514}{\nm} and intensity at \SI{1}{\milli\watt} were used. The sample was enclosed in a cylindrical chamber made of single-sided adhesive tape affixed to a plastic film (SecureSeal™-SA8R-2.5, diameter: \SIrange{8}{9}{\mm}, thickness: \SI{2.6}{\mm}, Grace Bio-Labs, Bend, OR) and sealed with a transparent cover to prevent evaporation.  Further details about the PISA can be found in the previous report\cite{Taniguchi2024-zs}.

To examine the correlation between actual object size and image size, non-fluorescent polystyrene beads of known diameters (\SIlist{22;99;300;600}{\nm}) were dispersed in ultrapure water and imaged.

\subsection{Particle tracking}
\label{extract_richpoor}
The movement of fluorescent beads dispersed within the suspension was monitored to assess the local viscoelastic properties of Laponite suspensions. To reduce the fluorescent beads' influence on structural formation, we minimized the concentration of beads to \SI{0.003}{\volpercent}. The beads were added to the filtered Laponite suspension before introducing TAMRA and NaCl. Fluorescence and particle tracking images were acquired as described above. A total of 500 images were captured at a fixed focal plane with a frame interval of \SI{0.43}{\second}. The fluorescent beads were excited at \SI{488}{\nm} and detected in the \SIrange{490}{540}{\nm} range. Bead trajectories were analyzed using Fiji’s Plugin, TrackMate\cite{Tinevez2017-td,Ershov2022-lt}.

To distinguish bead motion in different Laponite phases, \red{(i)} the red fluorescence channel (TAMRA, representing the Laponite structure) was averaged over 500 images and smoothed using a Gaussian filter ($\sigma = \SI{2}{\px}$) to generate a structural reference image. \red{(ii)} The mean fluorescence intensity along each bead trajectory, $\langle I(x,y)\rangle$, was calculated. \red{(iii)} The 30 trajectories with the highest intensity were classified as belonging to the ``Laponite-rich phase," and the 30 with the lowest as the ``Laponite-poor phase." Only trajectories tracked for more than 50 images (\SI{21.5}{\second}) were used for the analysis to exclude out-of-focus beads.

To quantify diffusion behavior, the mean squared displacement (MSD) was computed. The ensemble MSD was defined as follows:
\begin{equation}
    \mathrm{MSD}(\tau) = \frac{1}{N(t_{\mathrm{max},i}-\tau)}\sum_{i=1}^{N} \sum_{t=0}^{t_{\mathrm{max},i}-\tau} \left(\bm{r}_i(t+\tau)-\bm{r}_i(t)\right)^2
\end{equation}
where $\bm{r}_i(t)=(x_i(t),y_i(t))$ represents the position of the $i$-th bead at time $t$. The MSD was separately evaluated for the Laponite-rich and Laponite-poor phases. For the Laponite-poor phase, the MSD was fitted to $\mathrm{MSD} = 4D\tau$ using non-linear least squares (scipy, Python library) to determine the diffusion coefficient $D$.
\red{To evaluate viscoelastic properties, the complex modulus was obtained from the tracer bead MSDs using the generalized Stokes--Einstein relation\cite{Mason2000-ki}:  
\begin{equation}
    \left|G^*(\omega)\right| = \frac{k_{\mathrm{B}}T}{\pi a \, \mathrm{MSD}(1/\omega)\,\Gamma\left[1+\alpha(1/\omega)\right]} ,
\end{equation}
where $a$ is the tracer bead radius, $T$ is the absolute temperature, and $\Gamma$ denotes the gamma function.
The local logarithmic slope of the MSD $\alpha(\tau)$ was calculated from central finite differences of $\ln[\mathrm{MSD}(\tau)]$.  
The storage and loss moduli were then obtained as
\begin{equation}
    G^{\prime}(\omega) = \left|G^*(\omega)\right| \cos\left(\pi\alpha/2\right), \quad 
    G^{\prime\prime}(\omega) = \left|G^*(\omega)\right|
    \sin\left(\pi\alpha/2\right).
\end{equation}
This procedure was applied separately to bead trajectories in the Laponite-rich and -poor phases.}

To evaluate the heterogeneity in bead diffusion within a sample, we calculated the van Hove correlation function $P(\Delta x)$\red{, which is defined} as \red{the distribution of displacements along the $x$ direction} at a fixed delay time \SI{0.43}{\second} (corresponding to one frame). The displacement range was divided into 100 bins, and the probability was obtained by normalizing the counts in each bin. The function was computed separately for the Laponite-rich and Laponite-poor phases, and each was independently fitted to a Gaussian distribution.

\section{Results}

\subsection{Fluorescence observation of Laponite structures}

\begin{figure*}
    \centering
    \includegraphics[width=\linewidth]{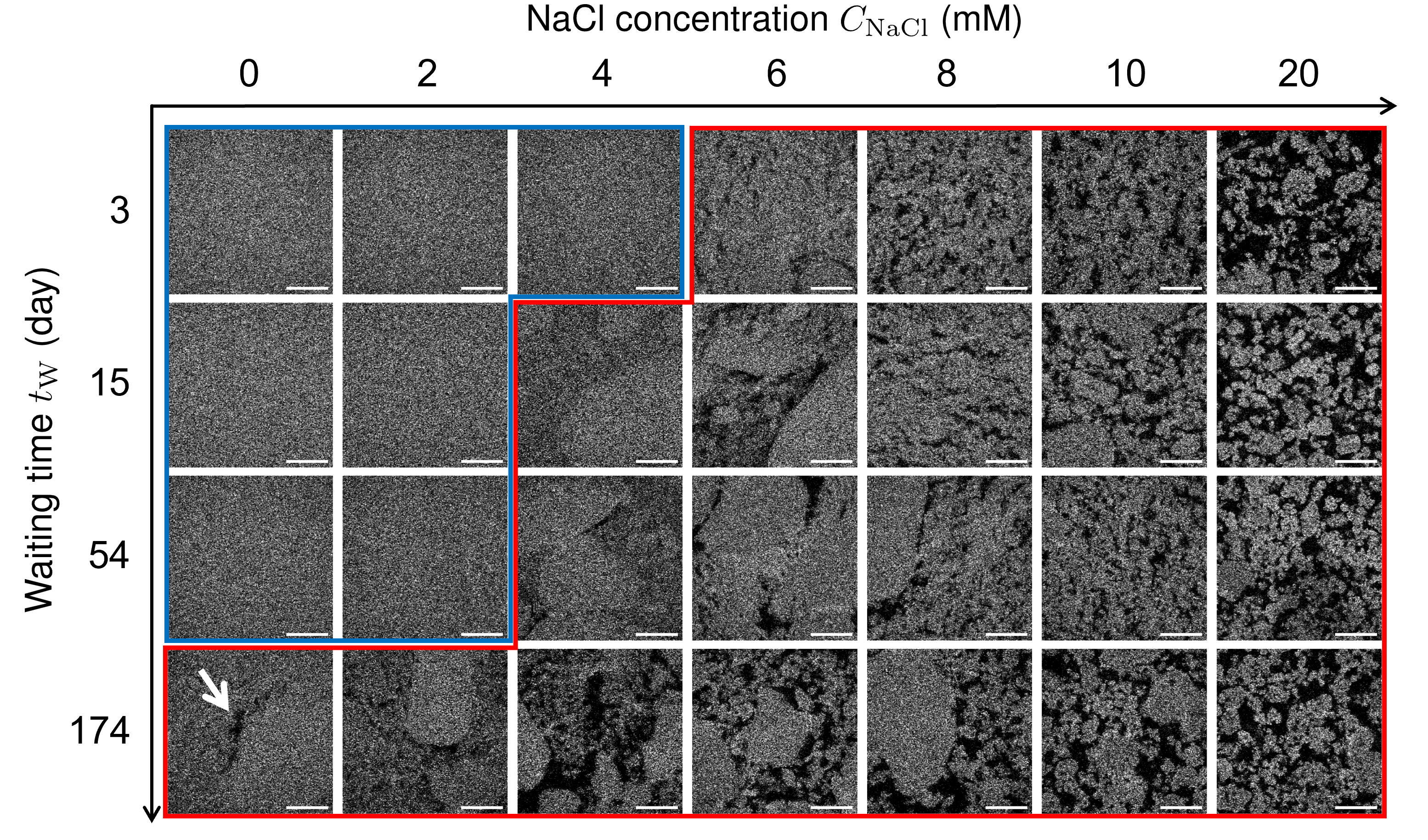}
    \caption{Confocal fluorescence images of \red{\SI{1}{\wtpercent}} Laponite suspensions at varying salt concentrations $\red{C_{\mathrm{NaCl}}}$ and waiting times $t_{\mathrm{W}}$.
    The bright areas indicate Laponite-rich regions.
    Images exhibiting homogeneous appearance are outlined in blue, while those exhibiting heterogeneous appearance are outlined in red.
    The arrow at $\red{C_{\mathrm{NaCl}}} = {}$\SI{0}{\mM}, $t_{\mathrm{W}} = {}$\SI{174}{\days}  indicates the crack. The scale bar represents \SI{50}{\um}. Images were acquired at a height of \SI{50}{\um} above the glass bottom to minimize the influence of the glass surface on structure formation. Fluorescence intensity is normalized by remapping the top and bottom 1\% of values to the upper and lower bounds, respectively.}
    \label{fig:fluor-label}
\end{figure*}

Confocal fluorescence microscopy was performed on fluorescently labeled Laponite samples to investigate the effect of salt concentration ($\red{C_{\mathrm{NaCl}}}$) on mesoscale structural changes associated with the aging of Laponite suspensions (characterized by the aging time, $t_{\mathrm{W}}$). As mensioned in Section \ref{fluor_observation_method}, the fluorescent dye TAMRA spontaneously adsorbs onto Laponite particles when added to the suspension. This adsorption was confirmed by measurements of translational diffusion of TAMRA (Section S1 and Fig. S1 in the Supplementary Information (SI)). The adsorption of fluorescent dyes to Laponite has also been previously reported for Rhodamine B, Rhodamine 3B, and Rhodamine 6G\cite{Garcia2012-wc,Pujala2018-ks,Lopez-Arbeloa1998-os}. The TAMRA concentration was kept at a minimal level sufficient for fluorescence imaging (\SI{20}{\nano\Molar}) to minimize any perturbation of the Laponite structure.

Fig. \ref{fig:fluor-label} illustrates representative confocal fluorescence images of Laponite suspensions at different salt concentrations $\red{C_{\mathrm{NaCl}}}$ and waiting times $t_{\mathrm{W}}$. These images reveal that the structural evolution of the suspensions is strongly dependent on salt concentration $\red{C_{\mathrm{NaCl}}}$ and varies with aging time $t_{\mathrm{W}}$.
At low salt concentrations ($\red{C_{\mathrm{NaCl}}} = {}$\SIlist{0;2}{\mM}), the systems initially exhibited a spatially homogeneous structure. After extended aging ($t_{\mathrm{W}} = {}$\SI{174}{\days}), cracks appeared at $\red{C_{\mathrm{NaCl}}} = {}$\SI{0}{\mM} (as indicated by an arrow), and a heterogeneous structure emerged at \SI{2}{\mM}.
These results indicate that at low salt concentrations, the aging dynamics of Laponite is extremely slow, maintaining a homogeneous structure over a long period. However, beyond \SI{174}{\days}, a gradual coarsening process may occur, eventually leading to the development of a heterogeneous structure.\par
As the salt concentration increased from \SI{2}{\mM} to \SI{20}{\mM}, pronounced structural inhomogeneities appeared at the early stage of aging, $t_{\mathrm{W}} ={}$\SI{3}{\days}. At moderate salt concentrations (\SIrange{4}{10}{\mM}), large aggregates of up to several tens of \si{\um} were observed throughout $t_{\mathrm{W}} = {}$\SI{54}{\days}. \red{For the shortest $t_{\mathrm{W}} = {}$ \SI{3}{\days}, the sample remained homogeneous at \SI{4}{mM}, but at higher salt concentrations, it became heterogeneous.} In contrast, at a high salt concentration ($\red{C_{\mathrm{NaCl}}} = {}$\SI{20}{\mM}), smaller aggregates formed early and remained unchanged in size, even after $t_{\mathrm{W}} = {}$\SI{174}{\days}. The difference in aggregate coarseness between moderate and high salt concentrations is more clearly visible in the corresponding images shown as a colormap (Fig. S2) and low-magnification fluorescence images (Fig. S3). These results suggest that coarsening is significantly suppressed as the salt concentration increases from moderate to high levels.\par

\red{To confirm our microscopic observations and rule out potential artifacts, we performed several control experiments. Without NaCl, the \SI{3}{\wtpercent} Laponite sample remained homogeneous (Fig. S4), similar to the \SI{1}{\wtpercent} sample. This indicates that the heterogeneity arises from the increased NaCl concentration, rather than from drying with a minor concentration increase. We also confirmed that the Laponite structure depends on the final NaCl concentration, rather than the NaCl concentration used for preparation (see Fig. S5 for the experiments using \SI{6}{\mM} and \SI{60}{\mM} NaCl solutions instead of \SI{1}{\Molar}). In terms of macroscopic behavior, the sample in the tube flows like a uniform solution when observing a homogeneous structure (low $C_{\mathrm{NaCl}}$, short $t_{\mathrm{W}}$) (Fig. S6, a). In contrast, when a heterogeneous structure is present (high $C_{\mathrm{NaCl}}$, long $t_{\mathrm{W}}$), the sample appears to exhibit a two-phase system, with a liquid phase on top and a gel-like phase at the bottom (Fig. S6, b).} \par
Overall, the fluorescence microscopy results reveal two distinct modes of salt-dependent aging. At low salt concentrations, the structure transitions from homogeneous to heterogeneous with increasing $\red{C_{\mathrm{NaCl}}}$, and the aggregate size evolves over time. In contrast, at high salt concentrations, submicron-scale inhomogeneities emerge rapidly and remain stable throughout the aging process.

\subsection{Power spectrum analysis of fluorescence images}

\begin{figure}
    \centering
    \includegraphics[width=0.8\linewidth]{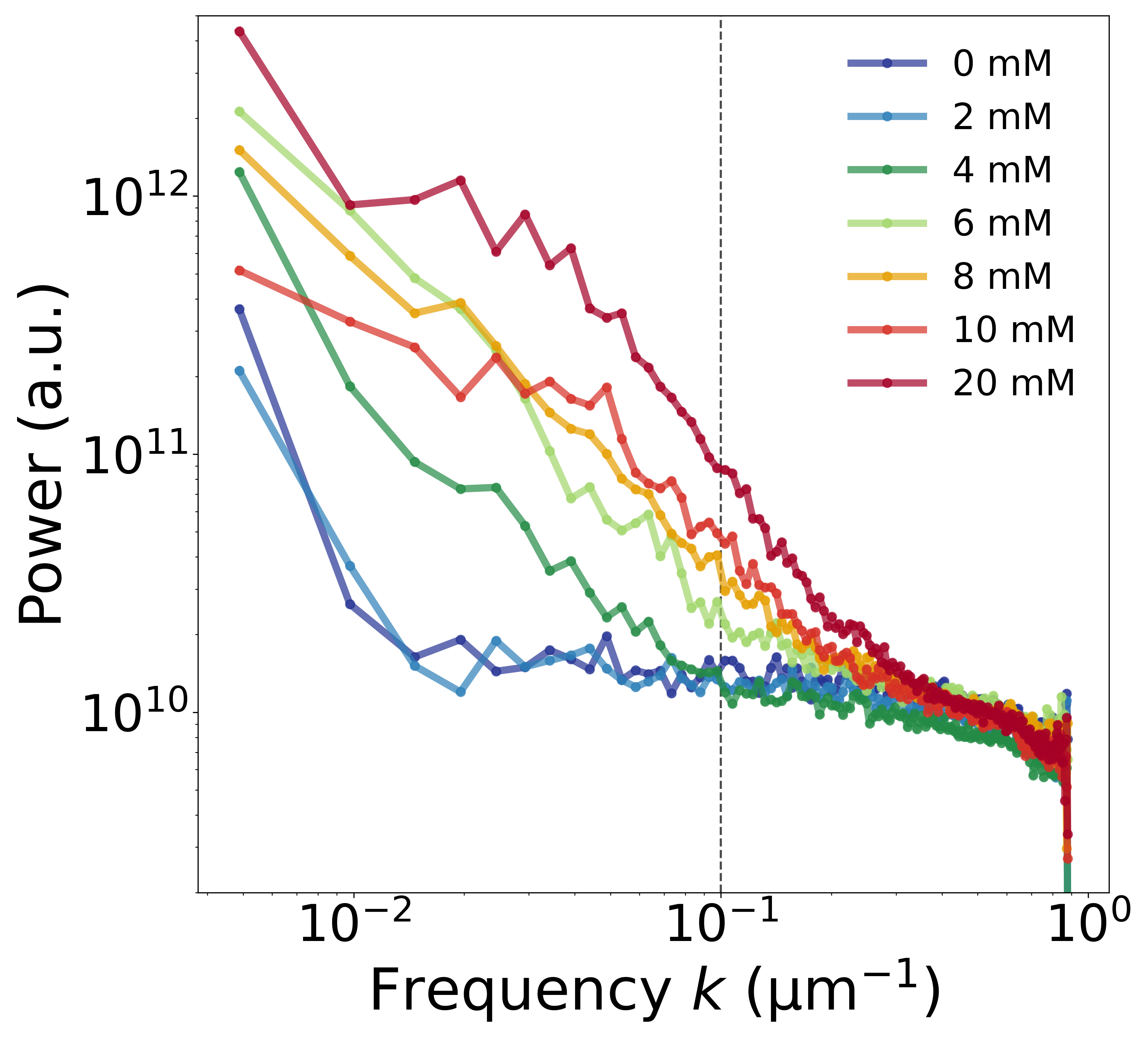}
    \caption{Power spectra of fluorescence images obtained from \red{\SI{1}{\wtpercent}} Laponite samples at different salt concentrations ($\red{C_{\mathrm{NaCl}}} = {}$\SIrange{0}{20}{\mM}) at a fixed waiting time of $\SI{54}{\days}$. Line colors correspond to different values of $\red{C_{\mathrm{NaCl}}}$.}
    \label{fig:ps_fluor}
\end{figure}

The power spectra of the fluorescence images, acquired at the same waiting time ($t_{\mathrm{W}} = {}$\SI{54}{\days}) stated in Fig. 3, were obtained to quantitatively assess the salt concentration dependence of the Laponite structure shown in the fluorescence images (Fig. \ref{fig:fluor-label}). At low salt concentrations ($\red{C_{\mathrm{NaCl}}} = {}$\SIlist{0;2}{\mM}), where the structure appeared uniform, the spectral intensity remained relatively constant in a wide frequency range, except at very low frequencies below $10^{-2}~\si{\um^{-1}}$. The increase in intensity in this low-frequency region might be attributed not to the Laponite structure itself but to experimental artifacts such as spatial gradients in fluorescence intensity caused by uneven illumination. 
In contrast, at moderate salt concentrations ($\red{C_{\mathrm{NaCl}}} = {}$\SIlist{4;6;8}{\mM}), where a coarse heterogeneous structure was observed, the spectral intensity was markedly elevated in the low-frequency region below $10^{-1}~\si{\um^{-1}}$. This result strongly suggests the presence of large aggregates with characteristic sizes on the order of tens of micrometers. Upon increasing the salt concentration to $\red{C_{\mathrm{NaCl}}} = {}$\SI{20}{\mM},  the frequency region showing elevated intensity extended to below $\sim2\times 10^{-1}~\si{\um^{-1}}$, indicating a reduction in the aggregate size.

These power spectrum analyses quantitatively support the microscopy observations: the Laponite structure transitions from a homogeneous state to a heterogeneous one featuring aggregates, and the characteristic size of these aggregates decreases from tens of micrometers to a few micrometers as salt concentration increases.

\subsection{Light scattering images of Laponite structures}

\begin{figure}
    \centering
    \includegraphics[width=\linewidth]{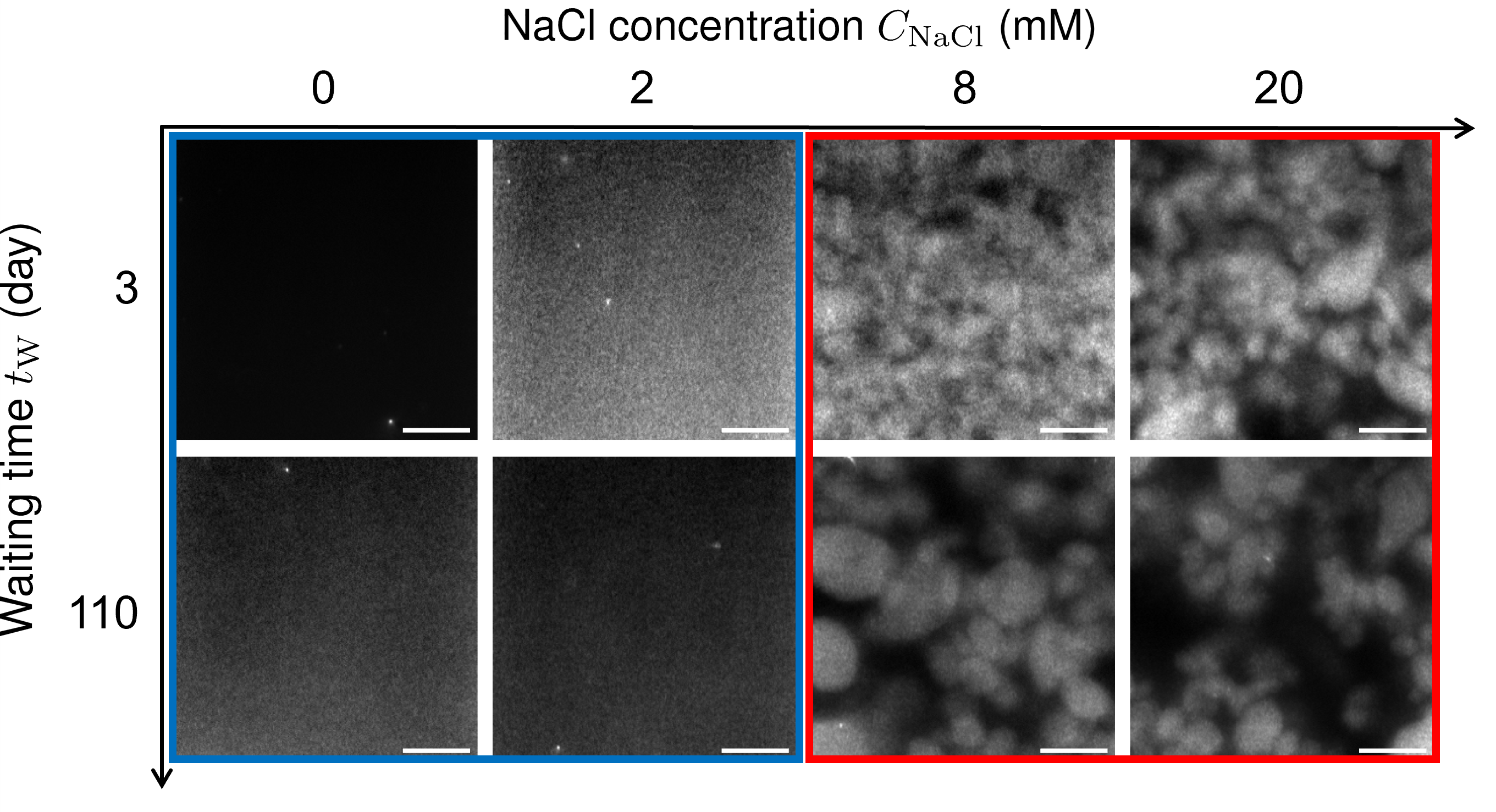}
    \caption{Scattering images of \red{\SI{1}{\wtpercent}} Laponite suspensions for various salt concentrations ($\red{C_{\mathrm{NaCl}}} = {}$\SIrange{0}{20}{\mM}) at a waiting time $t_{\mathrm{W}} = {}$\SIlist{3;110}{\days}.
    Blue and red outlines indicate homogeneous and heterogeneous structures, respectively, as in Fig. \ref{fig:fluor-label}.
    The scale bar represents \SI{50}{\um}.}
    \label{fig:scattering}
\end{figure}

To investigate the impact of fluorescent labeling on the structure formation of Laponite, we acquired images using label-free imaging techniques and compared these with fluorescent images (see Fig. \ref{fig:fluor-label}). As stated in Section \ref{light_scattering_method}, this method employs light scattering to visualize structures based on their inherent optical contrast. 
Isolated nanobeads (diameter $<{}$ \SI{1}{\um}) appeared as dots with a diameter of several micrometers, regardless of their size in the scattering images (see Section S2 and \red{Fig. S7} in SI). This occurs because the size of these beads falls within the submicrometer range, leading to diffraction-limited spots. Considering the uncertainties in the spatial intensity profile caused by diffusion and thermal drift, the spatial resolution of the imaging is estimated to be several micrometers. Therefore, the presence of spots measuring \SI{10}{\um} or more in the scattering images strongly indicates that the nanometer-sized Laponite particles are forming aggregates. 

Fig. \ref{fig:scattering} presents scattering images captured after waiting time, $t_{\mathrm{W}} = {}$\SIlist{3;110}{\days} and for samples with varying salt concentrations ($\red{C_{\mathrm{NaCl}}} = {}$\SIlist{0;2;8;20}{\mM}). At low salt concentrations ($\red{C_{\mathrm{NaCl}}} = {}$\SIlist{0;2}{\mM}), homogeneous structures are observed, while at higher salt concentrations ($\red{C_{\mathrm{NaCl}}} = {}$\SIlist{8;20}{\mM}), spatially heterogeneous aggregation structures emerge. These observations align with the trends in salt concentration and structural heterogeneity with submicon aggregates, observed in the fluorescence images (Fig. \ref{fig:fluor-label}). These findings indicate that fluorescence labeling has no significant impact on the Laponite structure, reinforcing the validity of using fluorescence observations for structural characterization.
%However, in terms of aggregate size, only subtle differences are observed between $\red{C_{\mathrm{NaCl}}} = 8$ and \SI{20}{mM}, making it difficult to determine any clear size trend from the scattering images alone.

\begin{comment}
\begin{itemize}
    \item Similar tendency to fluorescent labelling was observed
    \item TAMRA did not affect the aging of Laponite suspensions so much
\end{itemize}
\end{comment}

\subsection{Microrheology measurements in Laponite suspensions}

\begin{figure*}
    \centering
    \includegraphics[width=0.8\linewidth]{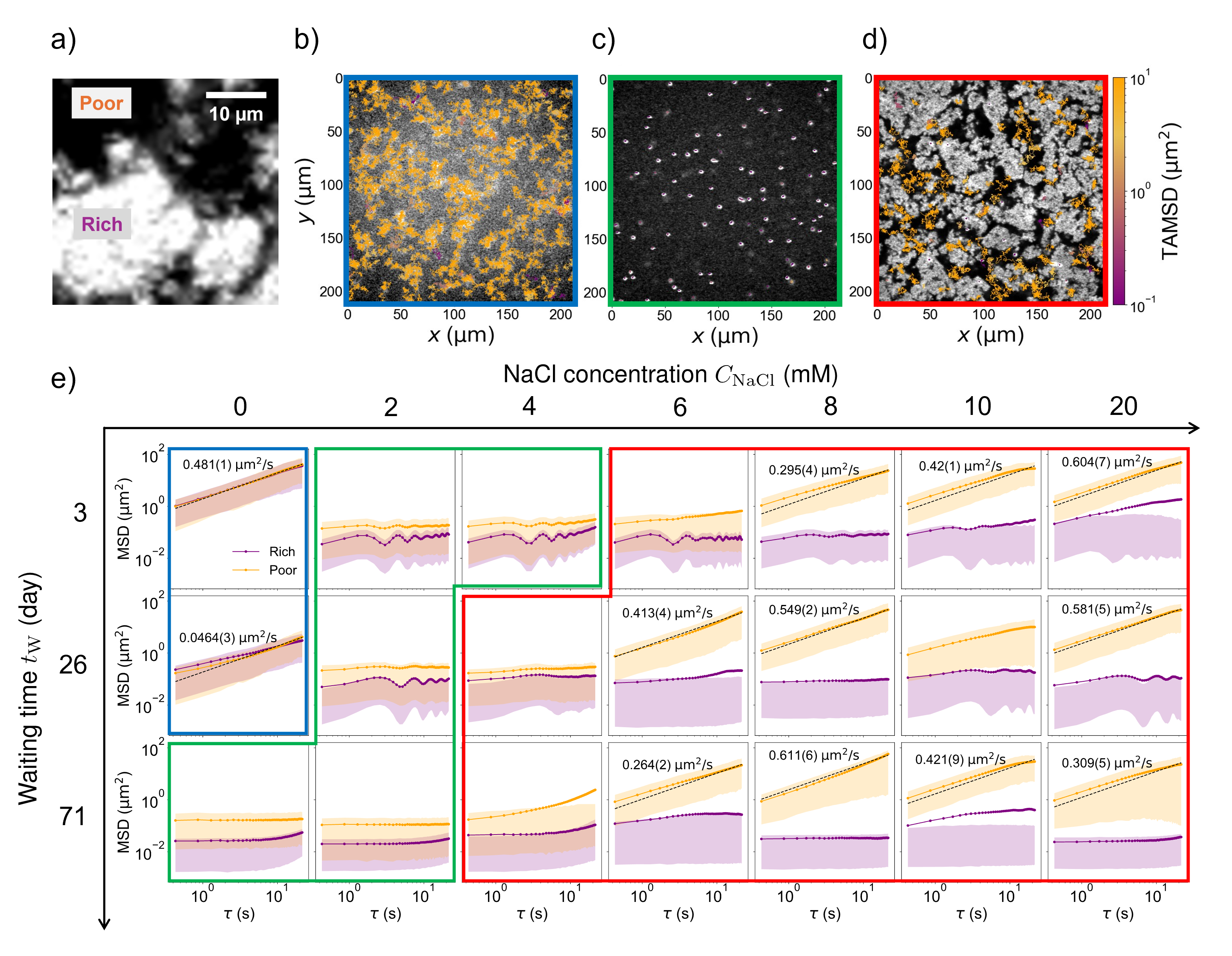}
    \caption{Translational diffusion of fluorescent beads dispersed in \red{\SI{1}{\wtpercent}} Laponite suspensions. (a) Representative time-averaged fluorescence image of a heterogeneous Laponite suspension ($\red{C_{\mathrm{NaCl}}} = {}$\SI{8}{mM}, $t_{\mathrm{W}} = {}$\SI{26}{days}), with Laponite-rich and Laponite-poor regions visible as high and low-intensity areas, respectively \red{(see Section 2.6(i); without Gaussian filtering)}. (b-d) Color-coded time-averaged squared displacement (TAMSD) trajectories show that the fluorescent beads exhibit three different diffusions depending on the Laponite structure \red{(outlined in blue, green, and red; see Fig. S9 and Section 2.6(ii) for the classification)}. (b) Normal Brownian diffusion in a homogeneous structure. (c) spatially uniform restricted diffusion in a homogeneous structure. (d) Heterogeneous diffusion in a heterogeneous structure.  (e) Ensemble-averaged mean squared displacements (MSDs) of 30 bead trajectories classified as diffusing predominantly through the Laponite-rich phase (purple) or the Laponite-poor phase (orange)\red{, with variability indicated by 16–84\% percentile bands ($n = 10^3 - 10^4$ for each curve) (see Section 2.6(iii) for further details)}. The black dashed line shows the fit to the poor-phase MSD using $\mathrm{MSD} = 4D\tau$, and the corresponding diffusion coefficient $D$ is also indicated with a fitting error. The numbers in parentheses represent the standard error ($1\sigma$) of the fitted diffusion coefficient, estimated from the fitting covariance assuming uncorrelated and normally distributed residuals.}
    \label{fig:overlay_msd}
\end{figure*}

The heterogeneous structure of the Laponite suspension strongly suggests the presence of spatial heterogeneity in its rheological properties. To explore the spatial correlation between structure and rheology, the diffusion behavior of fluorescent tracer beads was mapped onto structural images obtained by fluorescence microscopy.
To better elucidate the relationship between the Laponite microstructure and the mobility of the beads, the beads were classified into two groups based on their trajectories over a 215-second \red{(500-image)} observation period: those that predominantly diffused through the Laponite-rich phase and those that moved mainly through the Laponite-poor phase. Further methodological details are provided in Section \ref{extract_richpoor}.

The Laponite-rich and -poor regions were first identified by generating an average fluorescence image over a \red{500-image} interval, as illustrated in Fig.~\ref{fig:overlay_msd}a. The trajectories of the fluorescent beads were then superimposed on this image to produce composite overlays (Fig.~\ref{fig:overlay_msd} b–d), enabling the simultaneous visualization of structural features and bead dynamics.

In these overlay images, the bead trajectories are color-coded according to their time-averaged mean squared displacement (TAMSD), enabling visualization of the spatial relationship between local Laponite structure and bead mobility. The observed diffusion behavior can be broadly categorized into three types: normal Brownian diffusion in a homogeneous structure (Fig. \ref{fig:overlay_msd}b), spatially uniform restricted diffusion in a homogeneous structure (Fig. \ref{fig:overlay_msd}c), and spatially heterogeneous diffusion in structurally inhomogeneous regions (Fig. \ref{fig:overlay_msd}d).
Notably, Fig. \ref{fig:overlay_msd}d suggests that the large (orange) and small (purple) TAMSD trajectories align with phases of Laponite-poor and -rich regions, respectively. These data indicate that the observed spatially inhomogeneous diffusion of the beads strongly reflects the local structure.

\label{MSD_richpoor}

The spatial correlation between the structural inhomogeneity of Laponite and the corresponding heterogeneity in bead mobility is significant for understanding aging behavior. Following the method described in Section \ref{extract_richpoor}, we systematically selected 30 bead trajectories that primarily diffused through Laponite-rich regions (identified by high TAMRA fluorescence intensity) and 30 trajectories that mainly traversed Laponite-poor regions (low fluorescence intensity). \red{However, in the homogeneous sample, the intensity difference between the two regions is minimal, and the terms “rich” and “poor” are used only formally, without physical meaning, for readability and consistency in analysis.} This classification was applied consistently across all experimental conditions shown in Fig. \ref{fig:overlay_msd}b–d (see also Section S3 and \red{Fig. S8} in SI). Accordingly, we calculated the mean squared displacement (MSD) averaged across the ensemble using the selected 30 trajectories. In most cases, diffusion in the Laponite-poor phase is faster than in the Laponite-rich phase (Fig. \ref{fig:overlay_msd}e).

At low salt concentration ($\red{C_{\mathrm{NaCl}}} = {}$\SI{0}{\mM}), the diffusion behaviors in the Laponite-rich and Laponite-poor regions were found to \red{fall within 16–84\% percentile bands} at $t_{\mathrm{W}} = {}$\SIlist{3;26}{\days} (See MSD curves outlined in \red{blue} in Fig. \ref{fig:overlay_msd}e). This overlap indicates no significant difference in diffusivity between the two regions, consistent with the homogeneous structure observed via fluorescence microscopy. With extended aging ($t_{\mathrm{W}} = {}$\SI{71}{\days}), MSD became flattened, corresponding to spatially uniform restricted diffusion (Fig. \ref{fig:overlay_msd}c). This suggests that aging results in the development of a uniform gel-like structure at low salt concentrations.

As the salt concentration increases to moderate levels  ($\red{C_{\mathrm{NaCl}}}={}$\SI{6}{\mM}), the time at which flattened MSD curves appear becomes shorter, indicating an acceleration of the gelation process. Despite the apparently homogeneous structure suggested by fluorescence imaging, the MSD curves of beads in Laponite-rich and Laponite-poor regions do not  \red{overlap}, with the Laponite-poor phase showing slightly higher MSD values (See MSD curves outlined in green in Fig. \ref{fig:overlay_msd}e). This discrepancy may arise because slower (or faster) diffusing beads often appear brighter (or darker) in the structural images, which may affect the classification of the beads into the two categories.
\red{Some slow MSD curves ($C_{\mathrm{NaCl}} = {}$ \SI{2}{\mM}, $t_{\mathrm{W}} = {}$\SIlist{3;26}{\days}; $C_{\mathrm{NaCl}} = {}$\SI{4}{\mM}, $t_{\mathrm{W}} = {}$\SI{3}{\days}; $C_{\mathrm{NaCl}} = {}$\SI{6}{\mM}, $t_{\mathrm{W}} = {}$\SI{3}{\days}) exhibit minor oscillations ($\sim{}$\SI{0.1}{\um^2}), likely due to small fluctuations in fluorescence intensity, known as flickering.}

At high salt concentrations of $\red{C_{\mathrm{NaCl}}} \geq {}$\SI{8}{\mM}, or at $\red{C_{\mathrm{NaCl}}} = {}$\SI{6}{\mM} after extended aging ($t_{\mathrm{W}} \geq {}$\SI{26}{\days}), \red{MSDs from the Laponite-rich and -poor regions deviate beyond the 16–84\% percentile bands, demonstrating genuine dynamical differences} (see MSD curves outlined in red in Fig. \ref{fig:overlay_msd}e). In these cases, the MSD of the Laponite-poor phase exhibits a linear dependence on lag time, consistent with normal Brownian diffusion, similar to that observed at $\red{C_{\mathrm{NaCl}}} = {}$\SI{0}{\mM}. Therefore, The diffusion coefficient $D$ was extracted by fitting the MSD using the Brownian relation, $\mathrm{MSD} = 4D\tau$.

For the initial condition at $\red{C_{\mathrm{NaCl}}} = {}$\SI{0}{\mM}, the diffusion coefficient was determined to be $D \approx {}$\SI{0.481}{\um^2\per\second}.
In the Laponite-poor phase at high salt concentrations (Fig. \ref{fig:overlay_msd}e, outlined in red), 
$D$ typically ranged from \SIrange{0.2}{0.6}{\um^2\per.\second}.
Using these values along with the Boltzmann constant ($k_{\mathrm{B}} = {}$\SI{1.381e-23}{J/K}), bead radius ($a = {}$\SI{255}{\nm}), and ambient temperature ($T = {}$\SI{25}{\degreeCelsius}), the viscosity $\eta$ was estimated via the Stokes–Einstein relation: $\eta = k_{\mathrm{B}} T / (6\pi D a)$; 
this yielded $\eta \approx {}$\SI{1.78}{\milli\pascal\cdot\second} under the initial $\red{C_{\mathrm{NaCl}}} = {}$\SI{0}{\mM} condition and $\eta = {}$\SIrange{1.4}{4.3}{\milli\pascal\second} for the Laponite-poor phase at high salt concentrations.
The viscosity of the Laponite-poor phase at high $\red{C_{\mathrm{NaCl}}}$, which is only a few times greater than that of pure water (\SI{1}{\milli\pascal\cdot\second}), suggests that the Laponite-poor regions have not gelled, similar to the unaged Laponite suspension at low $\red{C_{\mathrm{NaCl}}}$. 

In contrast, the MSD curve for the Laponite-rich phase remains \red{essentially} flat and does not change over time, indicating that the Laponite-rich phase is in a gel-like, highly viscous region.
\red{However, at the shortest waiting times under high-salt conditions (e.g., $C_{\mathrm{NaCl}} = {}$\SI{20}{\mM} at $t_{\mathrm{W}} = {}$\SI{3}{\days}), the MSD curves in the rich phase are not completely flat. This likely reflects incomplete gel formation that allows some bead movement. These results suggest that, at high salt conditions, heterogeneous structures form rapidly, while the gel network in the rich phase strengthens gradually with time. We also estimated $G^{\prime}$ based on the generalized Stokes–Einstein relation. While fluctuations limited systematic analysis, the values obtained for $C_{\mathrm{NaCl}} = {}$\SIlist{0;20}{\mM} ($\sim{}$\SIrange{e-3}{e-4}{\pascal} and $\sim{}$\SIrange{e-1}{e-2}{\pascal}, respectively) agree with a previous report (see Fig. S10)\cite{Vyas2016-mf}. In addition, $G^{\prime}$ in the \SI{20}{\mM} rich phase increased with aging, supporting the view that the gel phase develops gradually over time.}

\subsection{van Hove correlation functions of bead diffusions}

\begin{figure*}
    \centering
    \includegraphics[width=0.8\linewidth]{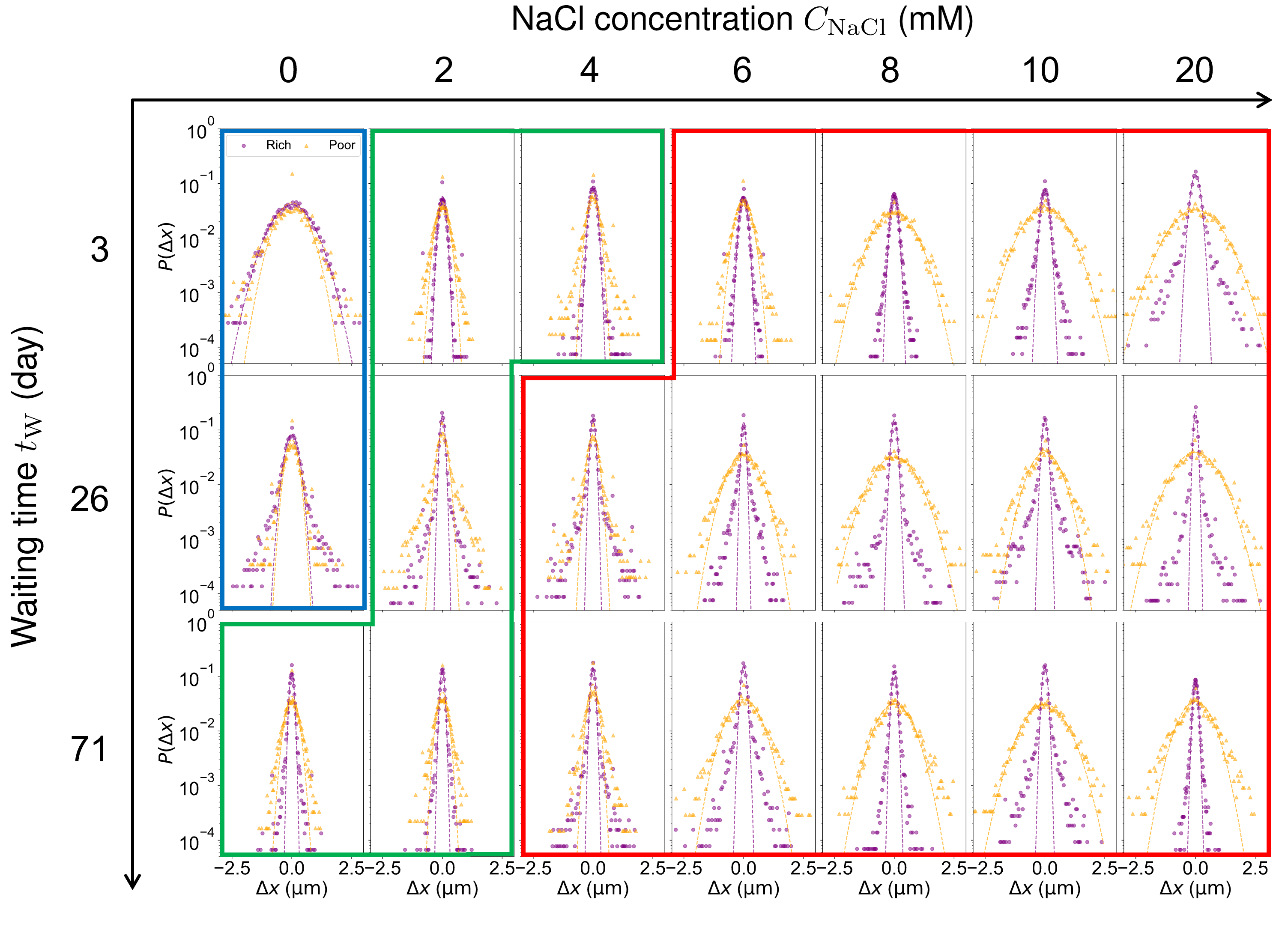}
    \caption{Van Hove correlation functions (VHFs)\red{, which is the distribution of displacements along the $x$ direction at a fixed delay time \SI{0.43}{\second}, in \SI{1}{\wtpercent} Laponite suspensions}. The trajectories of the beads used for the analysis are the same as those shown in Fig. \ref{fig:overlay_msd}. The three groupings (outlined by three colors (blue, green, and red)) are the same as in Fig. \ref{fig:overlay_msd}. 
    The dotted lines are Gaussian fits to the data.}
    \label{fig:vanHove}
\end{figure*}

To evaluate the structural heterogeneity on a spatial scale finer than that obtained from the MSD analysis, we independently calculated the van Hove correlation functions (VHF) for both the Laponite-rich and Laponite-poor regions (Fig. \ref{fig:vanHove}). The analytical method is detailed in Section \ref{extract_richpoor}. At low concentrations ($\red{C_{\mathrm{NaCl}}} = $\SIrange{0}{4}{\mM}), where the homogeneous structure was confirmed by fluorescence microscopy and MSD analysis, the VHFs of the Laponite-rich and Laponite-poor regions showed significant overlap. In contrast, at high salt conditions ($\red{C_{\mathrm{NaCl}}} \geq {}$ \SI{6}{mM}), the VHF in the Laponite-poor region is distinct from that in the Laponite-rich region. It shows a Gaussian distribution over a wide $\Delta x$ range. The conditions under which the Laponite-poor region shows a Gaussian distribution are almost the same as the conditions under which the MSD of the Laponite-poor phase usually shows normal Brownian diffusion (see the MSD curves outlined in \red{blue} in Fig. \ref{fig:overlay_msd}e). This Gaussian distribution suggests that bead diffusion in the Laponite-poor phase at high $\red{C_{\mathrm{NaCl}}}$ is approximately spatially uniform, similar to diffusion at low $\red{C_{\mathrm{NaCl}}}$.

The VHF profiles of the Laponite-rich regions (in high salt conditions outlined in red in Fig. \ref{fig:overlay_msd}e) show a sharp peak near $\Delta x = 0$ regardless of $\red{C_{\mathrm{NaCl}}}$ or $t_{\mathrm{W}}$, reflecting the high viscosity of the gel phase. Furthermore, the exponential-like tails deviating from the Gaussian distribution at large displacements strongly suggest the heterogeneous structure. Even at a low salt level (\SIlist{0;2}{\mM}), the VHFs of aged samples ($t_{\mathrm{W}} = {}$\SIlist{26;71}{\days}) also exhibit such tails, suggesting that microscale structural heterogeneity increases with aging.
Similar tails in VHFs were observed under slightly different conditions (salt, bead size, waiting time), but with the same Laponite concentration\cite{Oppong2008-td,Rich2011-uj} Given that the fluorescent beads used in this study have a diameter of \SI{510}{\nm}, this implies that the solid-like rich phase contains heterogeneous structures at submicron or nanometer scales.

This interpretation is consistent with the nanometric size of Laponite particles, which are approximately \SI{30}{\nm} in diameter. While the spatial resolution of our fluorescence imaging is limited to a few micrometers, it is reasonable to assume that Laponite particles locally aggregate to form mesoscopic structures below the resolution limit. 
Previously suggested structures, such as T-bonded configurations\cite{Ruzicka2011-zl}, further support the presence of submicron structures within the rich phase.

\section{Discussion}

The salt concentration dependence of heterogeneous structures in Laponite suspensions can be interpreted in terms of nonequilibrium gelation. Zaccarelli \cite{Zaccarelli2007-rs} proposed the concept of nonequilibrium gels, or arrested phase separation, in colloidal systems. In this framework, a colloidal suspension quenched to low temperature undergoes phase separation; however, the process is interrupted when the dense (rich) phase forms a glass, resulting in a percolated gel network.

In this study, we reinterpret the temperature axis of this phase diagram as the inverse of salt concentration. This is based on the assumption that higher salt concentrations suppress electrostatic repulsion and enhance attractive interactions between Laponite particles, thereby promoting phase separation. Thus, the addition of salt acts as a quench, with higher concentrations corresponding to deeper quenches.

At intermediate salt concentrations, the system enters the phase separation regime, and spatial heterogeneity gradually develops. However, the growth of the dense domains is hindered as the rich phase becomes dynamically arrested due to increasing viscosity. At high salt concentrations, strong interparticle attractions facilitate the formation of an attractive \red{gel} even at relatively low local densities. Consequently, phase separation is arrested at an earlier stage, and the resulting structure remains finely aggregated over a long time. This interpretation is consistent with the findings of Jabbari-Farouji et al. \cite{Jabbari-Farouji2008-pt}, who suggested that the rich phase becomes \red{dynamically arrested} under high-salt conditions. Supporting this view, molecular dynamics simulations by Oku et al. \cite{Oku2020-mi} showed that under deep quenches, the correlation length of glassy domains grows only logarithmically with time; this explains why, at \SI{20}{\mM} salt, our samples exhibited virtually no coarsening even after \SI{174}{\days}.
A related dependence on quench depth has been observed in oil-in-water nanoemulsions and other systems, where deeper thermal quenches lead to finer structures, as reported by Gao et al.\cite{Gao2015-mb} (see also refs.~12, 15, 22, 26, and 73 therein).

In contrast, at low salt concentrations (\SIlist{0;2}{\mM}), the suspensions exhibited mesoscopic homogeneity for over 50 days (Fig.~\ref{fig:fluor-label}). Ruzicka et al. observed similar behavior in salt-free Laponite suspensions under comparable conditions, specifically at low concentrations ($\le{}$\SI{1}{\wtpercent}), where Laponite suspensions initially formed a homogeneous gel-like phase that subsequently underwent macroscopic phase separation over several years\cite{Ruzicka2011-fq}. Rheological measurements of salt-free Laponite suspensions have indicated the development of attractive gels, rather than repulsive glasses\cite{Joshi2024-jx, Cocard2000-bj}.
Therefore, the homogeneous structure observed over an extended period is likely an attractive gel with nanoscale heterogeneity, whose meso-scale uniformity is maintained by strong electrostatic repulsion that delays heterogenization.
Furthermore, under the experimental conditions of this study, meso-scale heterogeneous structures emerged roughly after 75 days. It remains uncertain whether the onset of heterogeneity was spontaneous (for instance, a result of the gradual redistribution of salt ions during quiescent storage) or an unintended consequence of structural disruption (for example, through pipetting to position the sample inside the observation cell)\red{, as well as possible effects of interfacial ordering at the air–water interface}. Future experiments, such as monitoring completely sealed samples for several months \red{or using covered interfaces}, are expected to address these possibilities. \red{Further analyses, such as determining the fractal dimension, may also help us better understand the structural evolution.}

This scenario qualitatively accounts for the observed dependence of structural heterogeneity on salt concentration. Nevertheless, the precise boundaries of the phase separation region and glass transition line in the Laponite system remain unclear. Further studies are needed to construct a unified phase diagram, particularly in the low-concentration regime.

\section{Conclusion}

\begin{figure}
    \centering
    \includegraphics[width=\linewidth]{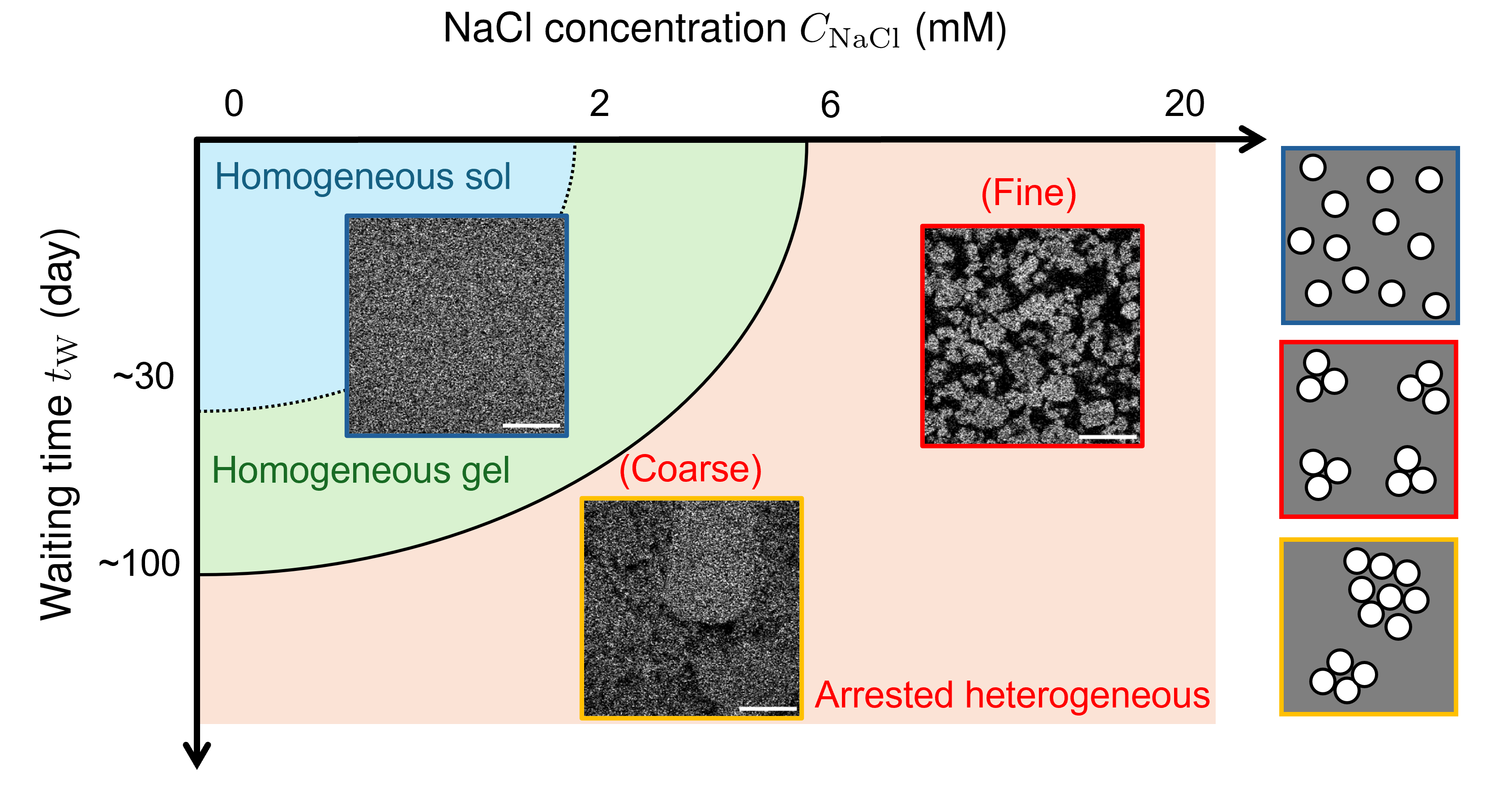}
    \caption{
    \red{Summary of the aging behavior of \SI{1}{\wtpercent} Laponite suspensions, highlighting phase behavior as a function of NaCl concentration and waiting time. Three regimes are identified: homogeneous sol, homogeneous gel, and arrested heterogeneous states (coarse and fine structures), illustrated with fluorescence images and schematic representations.}}
    \label{fig:summary}
\end{figure}

We investigated the aging-induced structural evolution of low-concentration (\SI{1}{\wtpercent}) Laponite suspensions with varying salt concentrations using fluorescence microscopy, label-free scattering imaging, and particle tracking microrheology \red{(Fig. \ref{fig:summary})}. Structural heterogeneity emerged more rapidly at higher salt concentrations (above \SI{6}{\mM}), with the characteristic size of aggregates decreasing as salt concentration increased. Fourier transform analysis quantified this trend, and the agreement between fluorescence and scattering images confirmed a negligible influence from fluorescent labeling. Particle tracking revealed distinct dynamics in Laponite-rich and Laponite-poor regions: the poor phase exhibited liquid-like behavior, while the rich phase showed gel-like characteristics. Nanoscale heterogeneities within the rich phase, inferred from the van Hove correlation function, likely originate from random network formation by the Laponite particles. These real-space observations provide insight into aging-induced structural heterogeneity in charged colloidal systems. The findings advance our understanding of complex phase behaviors of charged colloids, such as phase separation, gelation, and glassy dynamics, and support the development of colloidal gels for numerous applications.

%TC:ignore
\section*{Author contributions}
S. S. and M. Y. designed the research, and S. S. performed the fluorescence microscopy and colloidal diffusion experiments under the supervision of M. Y. Scattering imaging was conducted jointly by S. S., S. K., and Y. T. Data analysis and interpretation were performed by S. S and M. Y.. The manuscript was written by S. S. and M. Y., with critical comments by S. K. and Y. T.. All authors read and approved the final manuscript.

\section*{Conflicts of interest}
There are no conflicts to declare.

\section*{Data availability}
The data that support the findings of this study are available from the corresponding author, S. S. and M. Y., upon reasonable request.

\section*{Acknowledgements}

We thank Atsushi Ikeda and Hajime Tanaka for insightful discussions on the interpretation of the results.
This research was partially funded by the Japan Society for the Promotion of Science (JSPS) KAKENHI (grant numbers 22H01188, 24H02287 (M. Y.), 22K14800 (S. K), 19H05545 (Y. T.)) and the Japan Science and Technology Agency (JST) (grant numbers FOREST, JPMJFR213Y; CREST (JP- MJCR22E1) (M. Y.); ACT-X (JP-MJAX1914) (S. K), CREST (JP-MJCR2334) (Y. T.)), the Forefront Physics and Mathematics Program to Drive Transformation (FoPM) at the University of Tokyo (S. S.), and the Sasakawa Scientific Research Grant from The Japan Science Society (S. S.).

%%%END OF MAIN TEXT%%%


\balance


%%%REFERENCES%%%
\bibliography{rsc} %You need to replace "rsc" on this line with the name of your .bib file
\bibliographystyle{rsc} %the RSC's .bst file
%TC:endignore
\end{document}